# Monitoring dynamic networks: a simulation-based strategy for comparing monitoring methods and a comparative study

Lisha Yu[a*], Inez M. Zwetsloot[b], Nathaniel T. Stevens[c], James D. Wilson[d], Kwok Leung Tsui[a]

[a] School of Data Science, City University of Hong Kong, Hong Kong

[b] Department of Systems Engineering and Engineering Management, City University of Hong Kong, Hong Kong

[c] Department of Statistics and Actuarial Science, University of Waterloo, Canada

[d] Department of Mathematics and Statistics, University of San Francisco, U.S.A

**Abstract.** Recently there has been a lot of interest in monitoring and identifying changes in dynamic networks, which has led to the development of a variety of monitoring methods. Unfortunately, these methods have not been systematically compared; moreover, new methods are often designed for a specialized use case. In light of this, we propose the use of simulation to compare the performance of network monitoring methods over a variety of dynamic network changes. Using our family of simulated dynamic networks, we compare the performance of several state-of-the-art social network monitoring methods in the literature. We compare their performance over a variety of types of change; we consider both increases in communication levels, node propensity change as well as changes in community structure. We show that there does not exist one method that is uniformly superior to the others; the best method depends on the context and the type of change one wishes to detect. As such, we conclude that a variety of methods is needed for network monitoring and that it is important to understand in which scenarios a given method is appropriate.

## 1. Introduction

Dynamic network monitoring, also referred to as network surveillance, is becoming an active and interdisciplinary stream of research combining both network science and industrial statistics.

Generally speaking, network monitoring is focused on detecting abnormal behavior, such as a transient change or a persistent change in a network.

Complex systems can be modeled via networks where components of the system are treated as nodes and interactions among them are described by edges. Often the structure of such networks change over time and are thus considered *dynamic*. Examples of dynamic networks include social networks such as a Facebook or LinkedIn network, or an e-mail system where individuals send and receive e-mails. Other examples include biological networks that, for example, model communication between regions of the human brain, airline networks that describe the system of connections between different airports, or computer networks where individual computers connect to a system of servers. In each case, the structure of the network depends on the moment in time at which it is observed.

Interest in dynamic network monitoring is driven by the desire to detect changes in these systems (networks). Such a change from the normal baseline level of interaction may be an indication of an interesting change in the structure of the network or the behavior of the individuals that compose it. The main objective of network surveillance is to prospectively monitor a dynamic network and identify as quickly as possible when a change has occurred. Recently, network surveillance has become a burgeoning research area, though the field is relatively young and there is still a large opportunity for more additional work. For contemporary overviews of the field see Woodall et al. (2017) and Jeske et al. (2018).

Most existing methods for network monitoring are designed for a specific application or network configuration, and when a new method is proposed, it is typically clear in which situations the method is appropriate. However, the circumstances in which the method's use is inappropriate are often overlooked. As such, the broader applicability of such methods to other network surveillance problems is often difficult to determine, and at present it is not clear when to use a particular method and when not to. For example, Priebe et al. (2005) proposed a scan-based monitoring method and only evaluated the method by illustration with a case study. However, as pointed out by Jeske et al. (2018) "as new surveillance strategies become available, it is important to compare their performance with existing strategies…demonstration on specific examples…is not sufficient." More recently, computer simulations using synthetic networks have been proposed as a means to evaluate a method's performance as they facilitate comparison to a known truth.

Recent work featuring performance evaluations via simulation include Wilson et al. (2019), Hosseini and Noorossana (2017) and Yu et al. (2018).

However, what is still lacking in such investigations is a thorough comparison across a larger number of competing methods. As we illustrate in this paper, the design of simulation-based performance evaluations and comparisons is complex, and many choices need to be made. As such, it is unsurprising that most existing studies do not compare network monitoring methods thoroughly.

Thus, in this article, we develop a standard simulation-based approach for comparing the performance of network monitoring methods. We illustrate this method by comparing various network monitoring methods. In the next section, we propose a general framework for such investigations, and in Section 3, we describe the specific simulation study we use to compare several existing social network monitoring methods. In Section 4, we describe the results of that simulation and Section 5 concludes the article with a summary and final remarks.

## 2. A simulation-based strategy to compare network monitoring methods

The typical strategy for assessing the performance of a network monitoring method is to apply the method to a well-studied network application that best fits the criteria of the method itself (Savage et al. 2014; Woodall et al. 2017; Zhao et al. 2018). Without a formal comparison to other methods, the practitioner is lead to believe that every method is the "best" method. Simulated networks are required to fully evaluate the performance of a monitoring method. Simulated networks allow us to understand not only how and when a particular method performs well, but also when a method does not perform well. Such an investigation requires several choices to be made. Because this investigation is conducted via computer simulation, the choices one must decide on the manner in which the data (i.e., the networks) are generated. One must next decide on the monitoring objective from which data is observed. The monitoring objective largely depends on the application. For example, in a dynamic biological network, one is often interested in identifying small local changes among the interactions of the biological unit. On the other hand, in social networks, the primary goal is to detect communication outbreaks among groups of individuals as well as large structural changes in the community structure of the individuals over time. In this paper, we use social networks as an example to further elaborate the general framework. It is important to emphasize that our proposed strategy is suitable for any type of network.

A family of simulated networks should reflect two primary characteristics of dynamic networks: a) the dependence within and between networks in the sequence, and b) the time and type of change in the network sequence. The dependence in the network sequence can be characterized by dynamic network generative models. One must decide, for instance, whether the sequence of simulated networks are independent and identically distributed, or whether a dependency is imposed. In the social network literature, several statistical models have been proposed for analyzing networks. Static network models characterize the observed set of links of a single network; whereas, dynamic network models control the mechanisms that govern changes in the network over time (Goldenberg et al., 2010). One can use dynamic models directly to generate a sequence of networks. Well-known models include continuous time Markov chain models (Holland and Leinhardt, 1977; Snijders, 2005; and Wasserman, 1980) and discrete time Markov models (Hanneke et al., 2010, and Sarkar and Moore, 2006). Alternatively, one can use the static models to generate the dynamic network by random sampling, see the works of Wilson et al. (2019) and Yu et al. (2018). There are many static models available; for example, the Erdös-Rényi model (Erdös and Rényi, 1960), exponential random graph models (Wasserman and Pattison, 1996), the degree corrected stochastic block model (Karren and Newman, 2011), and the latent space model (Hoff et al., 2002).

The second characteristic that we consider is the type of change in the selected generative model. Use social networks as an example, we model changes that are common in social networks with community structure, including the birth and death of a community, merging of two communities, the division of a community, and others. We discuss these in more detail in Section 2.2.

**2.1 Dependent dynamic network generation**

We are interested in generating an ordered sequence of networks that demonstrate various types of change. Let $G_t$ represents the observed network at time $t$. Like Wang et al. (2017), we assume each network $G_t$ is generated from some known underlying model that governs the generative process. That is, the configuration of the network at time $t$, $G_t$, depends on a generative model $M_t$, and the network configuration at time $t-1$, $G_{t-1}$. We denote the generative model as $M = <$ $Type, \Theta, \alpha >$, that is characterized by (i) $Type$, which specifies the network model, (ii) $\Theta$, which represents the network model parameters and (iii) $\alpha$, which is the continuity parameter. The continuity parameter $0 \leq \alpha \leq 1$ controls the proportion of edges that are retained from the previous network. A dependent network at time $t$ is assumed to be generated in the following way: for each edge, keep the connection from time $t-1$ with probability $1 - \alpha_t$, and with probability $\alpha_t$, generate the connection according to the network model $Type_t$. In simulating networks in this fashion, the networks in the sequence of generated networks are dependent (for $\alpha_t \neq 1$). Furthermore, networks generated from same generative model resemble each other although they are not identical due to randomness. Only when the generative model $M$ changes do we consider it a "network change". As such, a network change can arise from a change in the network model, the network model parameters or the continuity parameter. Formally, to generate a dynamic network with change, we generate an ordered sequence of random networks $\boldsymbol{G}(T) = \{G_1, \dots, G_T\}$ according to

$$G_t = \begin{cases} M_t, & t < t^* \\ M_t^*, & t \geq t^*. \end{cases} \qquad (1)$$

By simulating $\boldsymbol{G}(T)$ as in (1), we introduce a change in the network at time $t^*$ that persists across the remaining networks in the sequence. In this way, $\boldsymbol{G}_{typical} = \{\boldsymbol{G}_1, \dots, \boldsymbol{G}_{t^*-1}\}$ are generated as "typical" networks; whereas, $\boldsymbol{G}_{anomalous} = \{\boldsymbol{G}_{t^*}, \dots, \boldsymbol{G}_T\}$ are "anomalous" networks. The overall network generative process is illustrated in Figure 1. The goal of the monitoring method then is to "signal" as quickly as possible following the time point of change $t^*$.

[insert Figure 1 here]

### 2.2 Degree corrected stochastic block model

Although any network model (*Type*) may be used to generate network $\boldsymbol{G}$, in this paper we use the degree corrected stochastic block model (DCSBM). In the following, we describe this model and demonstrate various types of network changes induced by changes to the model parameters. In this paper, we only focus on the changes arising from the network model parameters, i.e., $M^* = < Type, \Theta^*, \alpha >$. It is possible to change the type of network model (*Type*) and the continuity parameter ($\alpha$). However, we limit our scope and assume they do not change over time.

We use the DCSBM because it captures two crucial features of (social) networks: (i) community structure and (ii) degree heterogeneity. The communities of a network refer to subnetworks for which the level of interaction between nodes within communities is larger than between communities. Let $G = ([n], E)$ be an undirected network with $n$ nodes, $k$ communities, and edges E. Let $\boldsymbol{c} = (c_1, \ldots, c_n)$ denote the community labels where $c_i$ is the community label for node $i$. The DCSBM accounts for community structure through a $k \times k$ symmetric matrix $\boldsymbol{P}$ where the element $P_{c_i, c_j}$ dictate the propensity of connection between nodes in communities $c_i$ and $c_j$. Note that the community label of each node is often assumed unknown, but in practice, an appropriately chosen community detection algorithm may be used to determine them (see Fortunato (2010) and Porter et al. (2009) for reviews). Degree heterogeneity refers to the variability in individuals' likelihood to interact with others, regardless of their community membership. The DCSBM accommodates this phenomenon with a degree parameter $\boldsymbol{\theta} = (\theta_1, \ldots, \theta_n)$ which allows for each node in the network to have a different propensity for connection. Hence for M we consider *Type* as a DCSBM in this paper with parameters $\Theta = (\boldsymbol{c}, \boldsymbol{P}, \boldsymbol{\theta})$, i.e., $M = <\text{DCSBM}, (\boldsymbol{c}, \boldsymbol{P}, \boldsymbol{\theta}), \alpha>$.

Mathematically, the network $G$ can be represented by its adjacency matrix, an $n \times n$ symmetric matrix $\boldsymbol{A} = [A_{ij}]$, where $A_{ij}$ is equal to the number of interactions between nodes $i \neq j$. Since self-loops are not allowed, the adjacency matrix has zeros on its diagonal. Given $\boldsymbol{c}, \boldsymbol{P}$ and $\boldsymbol{\theta}$, under the DCSBM, the edge variables $A_{ij}$ are independent Poisson random variables with mean $\theta_i \theta_j P_{c_i c_j}$, i.e.,

$$A_{ij} \sim Poisson\ (\theta_i \theta_j P_{c_i c_j}). \tag{2}$$

The parameter $\boldsymbol{\theta}$ is arbitrary to within a multiplicative constant, which is absorbed into the $\boldsymbol{P}$ parameter. Thus, the DCSBM is not identifiable without some constraints on $\boldsymbol{\theta}$. Here we require the sum of $\theta_i$´s in the same community to equal the number of nodes in that community, namely,

$$\sum_{i:C_i=r} \theta_i = n_r, \tag{3}$$

for all $r = 1,2,\ldots, k$, where $n_r$ denotes the number of nodes in community $r$. For a more detailed overview of the DCSBM, see the work of Karrer and Newman (2011).

We note that the DCSBM is a generalization of the stochastic block model of Nowicki and Snijders (2001). That is, when $\theta_i \equiv 1$ for all $[n]$, the resulting network is a realization of a stochastic block model. Moreover, when $\theta_i \equiv 1$ and $P_{C_i,C_j} \equiv p \in (0,1)$ for all $c_i, c_j \in [k]$, then the DCSBM reduces to the Erdös-Rényi (ER) random graph model with probability parameter $p$ (Erdös and Rényi, 1960).

By using the DCSBM as the network model $Type$ and simulating a dynamic network $\boldsymbol{G}$ as in (1), we are able to simulate several types of changes by altering the parameters $\Theta = (\boldsymbol{c}, \boldsymbol{P}, \boldsymbol{\theta})$ that govern the DCSBM from time $t^* - 1$ to $t^*$. The changes $\boldsymbol{c} \to \boldsymbol{c}^*$, $\boldsymbol{P} \to \boldsymbol{P}^*$, and $\boldsymbol{\theta} \to \boldsymbol{\theta}^*$, each reflect a different type of change in the dynamic network, as described in Table 1.

[insert Table 1 here]

## 2.3 Network monitoring problem

The goal of network monitoring is to distinguish anomalous behavior from typical behavior in an ordered sequence of network observations $\boldsymbol{G}(T)$. A general framework for network monitoring is statistical process monitoring (SPM) (Wilson et al. 2019). The philosophy behind SPM is to distinguish between common-cause variation that is attributable to a relatively stable underlying process, and special-cause variation that is unusual relative to the underlying process. In general, SPM provides a methodology for the real-time monitoring of any characteristic of interest.

To perform network monitoring, one generally first specifies a statistic $S_t$, or more generally a vector of statistics $\boldsymbol{S}_t$, that provides some local or global summary of the network $G_t$. The choice of $S_t$ is flexible. Once a statistic $S_t$ has been chosen, SPM methodologies are used to distinguish anomalous behavior from typical behavior. It corresponds to the real-time identification of unusually larger or small values of $S_t$. The most popular technique used to determine the extremity of $S_t$ is a control chart with control limits that indicate boundaries of typical behavior. The monitoring process often consists of two phases, Phase I and Phase II. Phase I includes the retrospective analysis of baseline data to understand network process behavior. This may be a fixed period of time or a moving window. Control limits of $S_t$ are estimated in Phase I and used to design methods for prospective on-going monitoring in Phase II. In Phase II, when each new network is observed, we make a decision about the stability of the process relative to the Phase I baseline. The observed network $G_t$ is deemed typical if $S_t$ lies within the control limits and deemed anomalous if $S_t$ lies outside of these control limits.

Simulation is required for the systematic evaluation of a monitoring method's performance; only in such a controlled environment can an investigator be sure if and when a change in the network actually occurred, and whether the monitoring method identified the change in a timely manner. In this context performance metrics are required to evaluate a monitoring method's ability to detect network changes in simulated studies. The standard performance metric in SPM literature is what is referred to as the "run length" (RL) which is the number of networks observed before the monitoring method signals a change in the dynamic network. In particular, the Average Run Length (ARL) is often used as it characterizes the amount of time one might expect to wait for a change to be signaled. This value is desired to be large when the process is stable and low when the process change occurs. We note that McCulloh and Carley (2011) defined the average detection length metric, and Zhao et al. (2018) defined an average time-to-signal, which are both proportional to the ARL. The ARL metric is useful when a network change in the process is sustained until it is detected. However, if a change to the network is temporary, then more reasonable performance metrics are the conditional expected delay (CED) suggested by Shiryaev (1963) and the probability of successful detection (also called detection rate) suggested by Frisen (1992). The CED is the delay one can expect between the actual change time $t^*$ and the observed signal. The detection rate is the probability to detect a change in a given limited time period after it happened. Given a time limit $d \geq t^*$, the CED is denoted by CED = $E[RL | 1 \leq RL \leq d]$ and the detection rate is denoted by $P(1 \leq RL \leq d)$. Both metrics describe the ability to detect the change within a certain time. The ability to make a very quick detection (small $d$) is important in surveillance of sudden major changes, while the long-term detection ability (large detection rate) is more important for ongoing surveillance where smaller changes are expected. Ideally, a method has a large detection rate and a small CED, reflecting that the method has a high probability of detecting a change and also detects it quickly.

## 3. Comparative study

In this section, we illustrate the proposed comparison framework with a simulation study used to compare several network surveillance methodologies. As it is impossible to consider every existing monitoring method, we investigate three typical and representative classes of network monitoring methods that are chosen specifically to exemplify different areas of emphasis in social network monitoring applications. Specifically, we investigate a collection of nonparametric methods based on the monitoring of global network summary statistics, a scan-based method proposed by Priebe et al. (2005), as well as two model-based methods (Wilson et al., 2019 and Yu et al., 2018). Each of these is discussed below.

To compare the aforementioned monitoring methods, we evaluate their performance over a broad range of change-scenarios defined in the context of the undirected and non-attributed DCSBM, as described in Section 2.2. The scenarios we consider are chosen to cover a wide range of realistic network changes, and they serve to illustrate when the considered methods are appropriate and when they are not. We provide an overview of the three classes of methods in Section 3.1 and we describe the various change-scenarios in Section 3.2.

## 3.1 Description of monitoring methods

In our simulation, we consider and compare 15 monitoring methodologies. These methods can be categorized as nonparametric methods based on global network summary statistics, scan-based methods and model-based methods. Each class (and each method within a class) is described in the following subsections. Table 2 provides an overview of the methods considered.

[insert Table 2 here]

### 3.1.1 Methods based on global network summary statistics

A reasonably straightforward and intuitive approach for monitoring network change is to monitor summary statistics that describe the network over time. In a series of papers, McCulloh and Carley (2008a, 2008b, 2011) and McCulloh et al. (2008) use control charts such as the cumulative sum (CUSUM) and exponentially weighted moving average (EWMA) control charts to monitor global network summary statistics. In particular, global network summaries such as average betweenness and average closeness are used as input to these control charts. A recent comparative study by Hosseini et al. (2018) indicates that the EWMA chart performs better than the CUSUM chart in detecting outbreaks of increased communication in a small group of nodes in dynamic networks using a selection of global network summaries. For this reason, we focus exclusively on the EWMA approach in this paper, which we describe as follows. Let $S_t$ be the global network summary calculated at time $t$. Instead of monitoring the statistics $S_t$ directly, the EWMA chart is a time series plot of $E_t$, a smoothed version of $S_t$:

$$E_t = \lambda S_t + (1-\lambda) E_{t-1}, \tag{4}$$

where $0 < \lambda \leq 1$ is a smoothing parameter and $Z_0 = \hat{\mu}$ is commonly chosen as the starting point for the moving average. The steady-state control limits associated with this chart are given by

$$\hat{\mu} \pm 3\hat{\sigma}\sqrt{\frac{\lambda}{(2-\lambda)}}. \tag{5}$$

Many different network summary statistics have been developed over the years. The EWMA strategy is applicable to any such summary statistics $S_t$ that describes the structure of a network. Table 3 shows the summary statistics we select in this study, labelled according to the categorization of Newman (2018). These summary statistics are selected because they are commonly used in the literature and represent many potential measures available for monitoring. For example, the average closeness or maximum closeness over all nodes in the network helps to give insight into group cohesion. In the application of the EWMA chart, we choose $\alpha = 0.5$ based on the comparative conclusion drawn in Hosseini et al. (2018).

[insert Table 3 here]

### 3.1.2 Scan-based method

A variety of methods have been proposed in the literature that are broadly classified as scan-based methods. Scan-based methods are based on local statistics considering small subgroups of nodes. In a frequently cited paper, Priebe et al. (2005) propose a moving window-based scan method for the detection of a change in a sequence of networks. The goal of the method is to detect whether small regions of nodes have an unusually large number of connections (communications) amongst themselves as compared to their previous connections and that of others. They define a local statistic $O_{t,v}^k$ as the number of the edges in a $k^{th}$ order neighborhood around node $v$ at time $t$, where $k = 0$, 1, and 2. The neighborhood is defined as the subnetwork composed of the nodes that have a geodesic distance at most $k$ from $v$. Note that the degree of a node is the same as $O_{t,v}^0$. The moving window-based scan statistics are calculated using a two-step standardization procedure within moving windows of length 20. In the first step, for $20 < t < 40$, standardized statistics $O_{t,v}^{k*}$ are calculated for each of three neighborhood sizes ($k = 0,1,2$) for each node $v$ using $O_{t,v}^{k}$'s in the moving window ($1 \leq t \leq 20$):

$$O_{t,v}^{k*} = \frac{O_{t,v}^{k} - \mathrm{E}(O_{t:1\leq t \leq 20,v}^{k})}{\max(\mathrm{SD}(O_{t:1\leq t \leq 20,v}^{k}),1)} \quad (6)$$

A lower bound of 1 is used for the estimates of the standard deviation to avoid signals for relatively small changes in network behavior. In the second step, the statistic $T_t^k = \max_{v \in [n]}\{O_{t,v}^{k*}\}$ is defined as the maximum of the first standardized statistic over all nodes. For $t > 40$, a second standardization is performed on the $T_t^k$ for $k = 0,1,2$ and $21 \leq t \leq 40$:

$$T_t^{k*} = \frac{T_t^k - \mathrm{E}(T_{t:21\leq t \leq 40}^{k})}{\max(\mathrm{SD}(T_{t:21\leq t \leq 40}^{k}),1)} \quad (7)$$

Prospective monitoring is carried out by observing ($T_t^{0*}, T_t^{1*}, T_t^{2*}$) through time. An anomaly is signaled when $T_t^{k*} > 5$ for at least one of $k = 0,1,2$.

It is important to note that this scan method requires 40 network observations before monitoring can commence. The advantage of this moving window approach is that the network's expected behavior can adapt over time as the network evolves naturally. However, as the window moves along, undetected anomalies are incorporated into the baseline, making it nearly impossible to detect a significant change in the network if the change is not identified almost immediately.

### 3.1.3 Model-based methods

Recently, Wilson et al. (2019) and Yu et al. (2018) propose and investigate the use of a dynamic version of the DCSBM to model and monitor dynamic networks that undergo local and global structural changes and propensity change, respectively. Both methods apply SPM techniques to monitor estimates of the parameters associated with the DCSBM, and each is described below. Strictly speaking, Wilson et al. (2019) also proposed a means of monitoring nodal propensity, but the Yu et al. method has been shown to be superior for this purpose.

**Shewhart chart:** Wilson et al. (2019) propose a monitoring method that detects structural changes in a network by using control charts to directly monitor the unique entries of $\widehat{\boldsymbol{P}}_t$, the estimated propensity matrix associated with the DCSBM. For communities $r$ and $s$, the estimates are calculated as $\hat{P}_{r,s} = \frac{m_{r,s}}{n_r n_s}$, where $m_{r,s}$ is the total weight of edges between them. With $k$ communities, a total of $\binom{k}{2}$ Shewhart charts for each unique statistic $\hat{P}_{r,s}$ are monitored. Each chart signals a change if the monitored statistic lies outside the chart's control limits $\hat{\mu} \pm 3\hat{\sigma}$, where $\hat{\mu}$ is the sample mean of the $m$ phase I network observations and $\hat{\sigma}$ is the moving range estimate for the standard deviation of these $m$ observations given by

$$\hat{\sigma} = \frac{\sqrt{\pi}}{2(m-1)} \sum_{j=2}^{m} |\hat{P}_j - \hat{P}_{j-1}|. \tag{8}$$

**Multivariate $T^2$ chart:** Yu et al. (2018) propose a multivariate surveillance methodology to monitor node propensity changes. Their method applies the compositional $T^2$ chart on each community in a dynamic DCSBM. To apply this method, at each time point $t$ the node propensity estimates in community $r$ are calculated: $\hat{\boldsymbol{\theta}}_{(r)t}$. Let $\hat{\mathbf{z}}_{(r)t}$ be their corresponding *ilr* (isometric log ration) coordinates defined by

$$\hat{z}_i = \sqrt{\frac{i}{i+1}} \log \frac{\hat{\theta}_{i+1}}{\sqrt[i]{\prod_{j=1}^{i} \hat{\theta}_j}}. \tag{9}$$

Where $i$ is the node index. To check if the node propensity changes in community $r$, the following statistic is monitored:

$$T^2_{(r)t} = \left(\hat{\mathbf{z}}_{(r)t} - \hat{\boldsymbol{\mu}}_{\mathbf{z}_{(r)}}\right)' \hat{\boldsymbol{\Sigma}}^{-1}_{\mathbf{z}_{(r)}} \left(\hat{\mathbf{z}}_{(r)t} - \hat{\boldsymbol{\mu}}_{\mathbf{z}_{(r)}}\right). \tag{10}$$

Here, the mean vector $\boldsymbol{\mu}_{\mathbf{z}_{(r)}}$ of the log-ratio coordinates is estimated by the sample mean vector, and the in-control covariance matrix $\boldsymbol{\Sigma}_{\mathbf{z}_{(r)}}$ is estimated by the successive difference estimator using the $m$ network observations in Phase I. The monitoring statistic from Equation (10) is compared with an upper control limit (UCL) to see if the network has changed. A commonly used upper control limit for a prospective Phase II study is

$$UCL = \frac{(n_r-1)(m+1)(m-1)}{m^2 - m(n_r - 1)} F_{0.9}(n_r - 1, m - n_r + 1). \tag{11}$$

where $n_r$ is the number of nodes in community $r$ and $m$ is the Phase I sample size. Note that in order to compute (10) the inverse of the covariance matrix $\Sigma_{z(r)}$ must be calculated, and for large networks, this may be difficult to do.

To summarize, both methods seek to identify the change by monitoring DCSBM parameter estimates with an appropriately chosen control chart. Such model-based methods are advantageous because they identify more than *when* a network has changed; they also provide insight into both *how* and *where* the network changed depending on which control chart signals. However, the performance of such methods may deteriorate if the model is misspecified – a point we return to in Section 5.

### 3.2. Simulation settings

In this section, we describe how we apply the general comparison strategy proposed in Section 2 in the context of our investigation, and in doing so we illustrate how the proposed framework can be used in practice. As was described in Section 2, this strategy facilitates the comparison and evaluation of each monitoring method and the assessment of its ability to identify different types of network changes under controlled circumstances on synthetic networks. The simulated dynamic networks are generated according to the generative process described in Section 2.1. The DCSBM is chosen as the network model $Type$ to flexibly accommodate the changes described in Section 2.2. We assume the network model $Type$ and the continuity parameter $\alpha$ do not change over time and only focus on the changes associated with the DCSBM parameters.

We consider dynamic networks configured at each combination of the following simulation settings: continuity parameter $\alpha = 0.5$ or $1$, network size $n = 40$ or $100$ with $k = 1$ or $2$ communities. Note that if a dynamic network is generated with $\alpha = 1$, the network instances are independent. This is the most common simulation approach. See, for example, Azarnoush et al. (2016), Wilson et al. (2019), and Yu et al. (2018). However, by setting $\alpha < 1$, we impose a time-dependence within the generated dynamic network that more accurately reflects that natural evolution a dynamic network. By considering both $\alpha = 0.5$ and $\alpha = 1$ we evaluate the methods' performance in both scenarios. As noted, the choice of $\alpha = 0.5$ implies that at time $t$ fifty percent of generated edges are identical to the network at time $t-1$. By altering the number of nodes in the network ($n = 40$ versus $n = 100$), we investigate the impact of network-size on the performance of methods. Note that for a given $P$ and a growing network size the network will become denser. Therefore, to ensure comparable results, we adjust the values of the elements of $P$ for different network sizes to ensure the expected degree of any single node remained fixed at roughly 8 (Wilson et al. 2019 and Ridder et al. 2016). The degree parameters $\theta_i$ are randomly generated from the uniform random distribution $\theta_i \sim U(0.5, 1.5)$ and then scaled to meet the constraint in Equation 3. For fixed values of $\alpha$ and $n$ we alter the values of **P** and **θ** to impose various types of network change which we describe below.

One community DCSBM ($k = 1$)

We first consider a DCSBM network model with no community structure (the whole network forms one community). Thus the matrix **P** reduces to a single number $p$. It is often of interest to identify "connection outbreaks" among a group of nodes in the network. A connection outbreak corresponds to an increase in the average number of connections in the whole network or a subnetwork. *Global* and *local* changes are created with different specifications of $p \to p^*$. A *global change* means the entire network undergoes the outbreak and a *local change* means only part of the network undergoes the outbreak. For the *local change*, we choose the size of the anomalous subnetwork to be $n/5$. In addition, we are interested in detection of node *propensity changes,* these refer to shifts in the individual connection levels. In other words, we change the degree heterogeneity parameter, **θ** $\to$ **θ**$^*$.

Two-community DCSBM $k = 2$

In a two-community network, we consider 5 different types of change. We consider both communication outbreaks as well as changes to the community structure. In many cases, changes in the community structure will be reflected by changes in the parameters describing the DCSBM. That is, each of these 5 changes is produced by changing the component values of **P** $\to$ **P**$^*$. For a detailed discussion on the relationship between changing the community structure **c** $\to$ **c**$^*$ and its relationship to **P** $\to$ **P**$^*$, see Wilson et al. (2019). In the case of two communities, we assume that the baseline community labels **c** are known. We discuss the impact of determining community membership via a community detection algorithm in Section 4.5.

We refer to the 5 types of changes as *intensified communication*, community *splitting*, community *merging*, community *formation* and community *fragmentation* inspired by Peel and Clauset (2015). In the *intensified communication* scenario, a communication outbreak is imposed in a single community by specifying $P_{11} \to P_{11}^*$. The changes to community structure are defined as: *split*, when one community divides into two; *merge*, when two communities combine (the opposite of split); *form* when the network has one community and nodes around this community and then these nodes start to form a community; and *fragment*, when one of two communities breaks up (the opposite of form). All of these changes can signify an important transition in the system, and so it is important to be able to detect such changes.

Figure 2 summarizes the network changes we consider in this comparative study. The column titled *parameter change representation* explicitly states the chosen parameter values when the network has 40 nodes. The numerical values of $\boldsymbol{P}$ used in the case when $n = 100$ can be easily calculated by retaining the same expected degree of a single node when the network size is 40. For example, the parameter change in a *global change* for a one community DSCBM is $0.08 \to 0.1$ when $n = 100$.

We also suspect that the size of each community in a two-community network might affect the performance of the monitoring methods. To investigate this, we also study different node configurations. Let $\bar{N} = (n_1, n_2)$ denote the node distribution in a two-community network where $n_1$ and $n_2$ represent the number of nodes in each community. In this study, we considered $\bar{N} = (0.1, 0.9)n, (0.25, 0.75)n$ and $(0.5, 0.5)n$ to represent large, medium and no disparity in community sizes.

[insert Figure 2 here]

To summarize, each method is tested on data generated by (i) a dynamic one-community DCSBM configured by a combination of continuity parameter, network size and network change and (ii) a dynamic two-community DCSBM configured by a combination of continuity parameter, network size, network change, and discrepancy in community size.

For each simulation setting, we generate $m = 200$ in-control networks followed by networks that have undergone a network change. The first 200 networks are the baseline of Phase I samples. Then at time $t = 201$, we implement the network change and we simulate the anomalous networks until $t = 250$. This is done independently 1000 times, and each time all of the methods are evaluated in their ability to detect the change quickly. To summarize the performance of each method, we use the detection rate and CED defined in Section 2 as evaluation metrics. For a given method, the detection rate is computed as the proportion of time (in 1000 simulation runs) the method signals a change at least once in the time range $201 \leq t \leq 250$. Then, given a signal has occurred, the CED is computed as the average number of time anomalous networks observed before the signal. For computational efficiency, we set the time limit $d = 50$ and assess the ability of methods to detect a change within a 50 observations.

## 4. Results and discussion

We now present and discuss the results of our comparison study. In Section 4.1, we first show the performance of considered methods in a *no-change* scenario. We then summarize the results in a one-community DCSBM in Section 4.2. For a two-community DCSBM, since there are 5 scenarios for each of three node configurations, we first discuss the obtained results considering $\overline{N} = (0.5, 0.5)n$ in Section 4.3. Then we discuss separately the impact of different node configurations in Section 4.4. Besides, we discuss the impact of determining community membership via a community detection algorithm in Section 4.5. All of our findings are reiterated and concluded in Section 4.6.

The results are summarized largely with figures. The top panels of each figure show the CED and the bottom panels show the detection rate. Different colored lines and bars represent different network sizes of $n = 40$ and $n = 100$ and difference network dependence levels of $\alpha = 0.5$ and $\alpha = 1$. On the horizontal axis we show each of the fifteen considered methods.

### 4.1 In-control performance

We begin by showing the performance of the methods on a *no-change* (in-control) scenario. In this case, any signal is a false alarm and one would expect that all of the methods show similar false detection rates for a fair comparison. Figure 3 shows the *no-change* scenario in a one-community network ($p = 0.2$) and Figure 4 shows the *no-change* scenario for the two-community network with $\boldsymbol{P} = \begin{pmatrix} 0.3 & 0.1 \\ 0.1 & 0.3 \end{pmatrix}$. Both figures show that the methods perform similarly in terms of the detection rate. Given that we observe false alarms, the CEDs are approximately 30. For the independent networks ($\alpha = 1$, blue and purple), the false detection rates are low for all methods, except method 5, 6 and 13. Method 5 and 6 are based on the closeness statistics and method 13 is the scan method. This implies that compared to the other methods, monitoring based on closeness or Priebe's scan statistic gives more false alarms. When the structure of the dynamic network is time-dependent ($\alpha = 0.5$), the false detection rates grow for most of the methods. This is in line with results for monitoring autocorrelated variables (Bisgaard and Kulahci, 2005). For most of the methods, the false detection rate becomes alarmingly high. These charts will signal a change in a rate from 20% to 50% – even though there is no actual change. This can become a serious drawback for these methods, as temporal dependence is often observed in network streams. One possible solution could be to adjust the control limits to obtain a lower false alarm rate. However, this would require detailed knowledge of the network's time dependence structure. Note that method 15 does not signal any false alarms, therefore the CEDs are not reported in Figures 3 and 4.

[insert Figure 3 here]

[insert Figure 4 here]

**4.2 One community**

*Global change and local change*

The first types of change we study are the *global* and *local* increases in the communication level. In the *global* scenario, all nodes start to communicate more intensely. Any methods are expected to signal on this obvious change. Figure 5 displays the results. We see that most methods detect the change in nearly 100% of the simulation runs, except methods 5, 6, 11, 13 and 15. When considering the CED, methods 1, 3, 7, 9 and 14 work the best, where they all signal the change within 10 time points on average.

We see low detection rates for methods 5 and 6 which are both based on closeness. As such, it appears as though the closeness of the nodes does not capture the change quite as well as the other measures. Notice that the performance is especially poor when the network size is small, i.e., $n = 40$. The poor performance of method 11 which is based on the maximum of shortest path length is expected because the maximum value of the shortest path length of each node apparently does not change significantly in the *global* scenario. The poor performance of the scan method (13) may seem unexpected. It can be explained by its moving window property, where the undetected anomalous networks are incorporated into its baseline. The poor performance of Yu's method (15) is unsurprising as it is designed to detect individual propensity changes and the $\theta's$ are scaled so that their average is always 1.

[insert Figure 5 here]

Next, in Figure 6, it shows the comparison when there is a *local* change in a one-community network. It is clear that method 1 perform the best, with a detection rate above 60% when $n = 40$ and above 90% when $n = 100$. Methods 3, 7, 9, 10 and 14 also show good detection power. These methods have higher detection rates when $n = 100$ because in a larger network there are a larger number of nodes whose communication levels have increased. Conditional on a method detecting the change, we see little variability in the CED values. Methods 2, 4, 5, 6, 11, 12, 13, and 15 do not adequately identify this change as these all have detections rates below 25%. Overall, under our selected circumstance where a little increase on $p$ has been induced for the anomalous subnetwork, none of the compared methods show good detection power. However, we can see the comparability among the considered methods. The performance of each method will improve as the changed on $p$ increases for the anomalous subnetwork. This can be illustrated in the later *intensified communication* scenario, which can be considered as a related extension of the *local* change scenario for the methods based on global network statistics.

[insert Figure 6 here]

*Propensity change*

Figure 7 shows the comparison results for the *propensity* change where nodes become more heterogeneous in their connection propensities. It is clear that method 15 works the best in terms of detection rate and CED. This is unsurprising because this method is designed specifically to detect propensity changes. Methods 5, 6, 7 and 8 – based on closeness or a clustering coefficient – also have high detection rates, but the CED is larger than for method 15 and so it takes a longer time for them to detect the change.

[insert Figure 7 here]

### 4.3 Two known communities

In the following scenarios, we assume that the network contains two communities and the community label of each node is known. As such, methods 14 and 15 have a slight advantage as they incorporate the community information, unlike the other methods which are based on the global network summary statistics (methods 1-12) or on a scan statistics (method 13).

*Intensified communication*

First, we consider *intensified communication* where the communication level in one of the two communities increases, i.e., $p_{11}$. The results for this scenario are depicted in Figure 8. Several methods are very capable of detecting this change. Method 14 is designed to detect these types of changes and it performs the best as expected. Method 1, which is based on degree, shows nearly equivalent performance to method 14. We also see that the global methods based on clustering coefficients work well (methods 7 and 8), where the global clustering coefficient (7) works much better than the local cluster coefficient (8). Method 3 based on betweenness and method 9 based on diameter also detect all shifts but their CED is less desirable than that of either method 1 or method 14. In general, we see improved performance of all methods in a larger network size.

[insert Figure 8 here]

*Merge and split*

Figure 9 shows the results when two communities merge into one community. It is clear from the figure that method 14 outperforms all the others. The other methods, though some have high detection rates (methods 3 and 9), all of them have undesirably high CED values. Besides, methods 2, 4, 5, 6, 11, 13 and 15 have detection rates less than 25%, and thus they are not adequately able to detect this type of change. Note that their detection rates are similar to the ones observed in the *no-change* scenario.

We also study the *split* scenario (the opposite of *merge*). In this scenario, a single community splits into two separate communities. As with the merging of communities, method 14 is the best at detecting community splitting. The results are very similar to Figure 9, so we do not display them here. It is important to note that the *split* scenario is not the same as the *local* change scenario discussed previously because in this case we assume that there exist two communities to begin with.

[insert Figure 9 here]

*Fragment and form*

Figure 10 shows the results of the *fragment* scenario. In this scenario, we begin with two communities with comparable communication levels before the change. After the change, however, one community shows a higher intensity and the other community starts to disappear. As with all of the two-community scenarios, method 14 works well. Figure 10 demonstrates that methods 1,3,7,8, and 10 also perform well. Surprisingly, the scan method does not work very well in this case. This is most likely due to the moving window which incorporates anomalous behavior that is not immediately detected.

[insert Figure 10 here]

**4.4 Impact of node configuration**

In each of two-community scenarios discussed above, we also evaluate the impact of the relative sizes of the two communities. In particular, we investigate the impact of changing the size of the first community to be either 50%, 25% or 10% of the total network size.

We start with the *intensified communication* scenario, where one (the small) community becomes more active, see Figure 11. The performance of the compared methods based on the global network summary statistics all perform worse as the size of the intensified community decreases. This is unsurprising since a change to a small number of nodes in a small network represents a very small signal-to-noise ratio that will be difficult for a global statistic to discern. Method 14 appears to perform reasonably well in each case, and it is the only method that performs adequately in the 10% case.

[insert Figure 11 here]

Next, we consider the *split*, *merge*, *fragment* and *form* changes. We discuss these together as the relative performance of the considered methods is similar across the three node configuration cases. Figure 12 depicts the results for *fragment* change. Interestingly, with respect to the global network statistics, the effect of node configuration is opposite of the *intensified communication* change; as the size of the first community becomes smaller, methods based on the global network statistics perform better and show near 100% detection rates as well as low CED values, see panels (b) and (c) in Figure 12. This result is likely because the second and larger community also changes in these four scenarios whereas it does not change in *intensified communication* scenario. Hence, when $\bar{N} = (0.1, 0.9)n$ and $(0.25, 0.75)n$, these four types of change not only affect the first small community but also the second large community, and it leads the methods based on global network statistics to identify the change easily.

Although not shown here, for *split* and *merge*, methods 2, 5, 6 and 12 do not perform well when $\bar{N} = (0.1, 0.9)n$. The differential performance of these methods in *split*/*merge* relative to *fragment/form* arises likely because the changes to the ***P*** matrix are larger in *fragment/form* (0.3 to 0.1) cases than in *split*/*merge* (0.2 to 0.1) cases. We also point out that method 14 shows good performance for each node configurations, and methods 13 and 15 are not capable of detecting *split*, *merge*, *fragment* or *form* scenarios, irrespective of the node configuration.

[insert Figure 12 here]

**4.5 Two unknown communities**

In the previous discussion of results in a two-community network, we assume the community label is known. Here we discuss the case when the community label is unknown. When the community structure of the dynamic network is unknown, the community label must be estimated using Phase I networks so that the parameters of a block model can be calculated in Phase II as it is necessary for the model-based methods. The estimation of the community labels, commonly known as *community detection*. Many detection methods have been developed, see Fortunato (2010) and Porter et al. (2009) for reviews. For monitoring purposes, we suggest using the regularized spectral method from Qin and Rohe (2013) to establish the community labels. In follows, we briefly comment on the uncertainty associated with community labels impacts the performance of the model-based methods and show the performance of Wilson's method by applying the community detection algorithm.

Adopting the Phase I vs. Phase II framework of the SPM literature, we assume that the networks generated in Phase I are identically distributed. Consider the average network from Phase I, whose ($i,j$)th entry is the average edge weight between node $i$ and node $j$ over Phase I networks. If the expected value of this network has no identical rows, then spectral clustering of the graph will provide asymptotically consistent community label estimates. In practice, this suggests that if the number of Phase I network is large enough, we obtain consistent estimators for the community structure for the sequence of graphs before the time change $t^*$. This fact is a consequence of the main result presented in Han et al. (2015). Error in community label estimation will introduce error in the parameter estimates, and this error will increase for short Phase I windows. We therefore advise using as many Phase I networks as possible, but the choice depends on the judgement of the practitioner and the availability of data.

In this realistic situation, in which the community labels are unknown, the performance of model-based methods like methods 14 and 15 may be seriously impacted. From the results already discussed, we have found that method 15, even when the community labels are known, does not adequately detect any of the two-community changes outlined in Figure 2. As such, we investigate the effect of community detection on method 14 only, and specifically in the *merge* and *split* scenarios. The results are presented in Figure 13. We see that when the community labels are estimated (unknown), Wilson's method does an excellent job at identifying community *merge* but a very poor job at identifying community *split*. These results agree with the conclusions of Wilson et al. (2019) with respect to the expected performance of the method when community labels are unknown.

[insert Figure 13 here]

**4.6 Overview of results**

We have compared 15 monitoring methods by testing their performance on a variety of realistic changes that a dynamic network may undergo. Table 4 summarizes the performance of all of the methods. For each method (and each type of change) we categorize its performance as good, moderate or poor and denote this with the symbols ✓✓, ✓ and ✗, respectively. A method receives a designation of "good" for a given change if it has (near) 100% detection rate for both $n = 40$ and $n = 100$ and for $\alpha = 0.5$ and $\alpha = 1$ as well as a CED less than 10. A method receives a designation of "moderate" for a given change if it has at least 75% detection rate for all $n$ and $a$ as well as a CED between 10 to 30, and a method is designated as "poor" for a given change if the detection rate never exceeds 25% for all $n$ and $a$. The rightmost column of Table 4 indicates which method performs best for the respective scenario.

[insert Table 4 here]

By examining this table, it becomes clear that no single method works well in all change scenarios. Method 14 due to Wilson et al. (2019) works very well in all scenarios. Notable exceptions include the *local* scenario (for which none of the methods work well) and the *propensity* scenario. However, it should be noted that the version of Wilson's method applied here deliberately does not include their method for detecting changes in individual interaction propensities. Method 15 is designed especially for this type of change, and so it shows strong performance in this scenario. A practical strategy would be to use these two methods in tandem, thereby increasing the number of types of change one is able to detect.

The methods based on the global network summary statistics work well if the majority of the nodes in the network change in some way. Otherwise, these methods break down. As such, in practice we recommend that a global network statistic-based method only be used if one anticipate detecting changes that affect more than half of the network. Note that in general, the global statistics based on a maximum of a certain statistic tend to show worse performance than their counterpart based on the average statistic. This is because the average statistics capture the change in parts of the network better, whereas the maximum values can be unchanged when a part of the network changes. Due to this finding, we do not recommend the use of global network summary statistics based on maxima.

Overall, when considering the problem of network monitoring in practice and choosing a monitoring strategy, a practitioner must think carefully about the type of changes they wish to identify. That says, we reduce the number of possibilities by recommending that methods 2(average eigenvector centrality), 4(maximum betweenness), 5(average closeness), (maximum closeness), 8(average local clustering coefficient), 12(assortativity) and 13(scan method) be eliminated from consideration because they are consistently out-performed by other alternatives. Ultimately, the best method depends on the network structure and the type of change one is interested in detecting. To detect …

- ...*global* change in a one community network use method 14 proposed by Wilson et al. (2019) or use an EWMA chart based on average degree, average betweenness, global clustering coefficient or diameter.
- ... *propensity* change use method 15 proposed by Yu et al. (2018).
- ... any change in a two community with unbalanced node configuration, i.e., $\bar{N} = (0.1, 0.9)n$, use method 14 proposed by Wilson et al. (2019) or use an EWMA chart based on average degree, average betweenness, global clustering coefficient or average path length.
- ...any change in a two community with balanced node configuration, i.e., $\bar{N} = (0.5, 0.5)n$, use method 14 proposed by Wilson et al. (2019) or use an EWMA chart based on average degree, average betweenness or average path length.

We note that the scan method proposed by Priebe et al. (2005) performed poorly in our study. This is likely due to the fact that we assume the dynamic network is stable over time and thus the moving window associated with that method is of little benefit. As an extension to this work one might consider adding some volatility to the simulated dynamic network, in which case the performance of their method would likely improve.

## 5. Conclusion

Network surveillance is an exciting practical problem and a tremendously fruitful research area. Many monitoring methodologies exist already, and many more are currently being developed. We strongly encourage that, as new methods become available, their performance is systematically compared to the existing state-of-the-art; illustrating what is effective on a single dataset or some contrived application is simply not sufficient. In this article, we propose a general simulation-based framework for the systematic comparison of monitoring methods, and we apply this framework to several state-of-the-art monitoring methods. Our findings are perhaps not surprising – the best method depends on which types of change one wishes to detect. That says, a hybrid of the Wilson et al. (2019) and Yu et al. (2018) methods seems like a promising general strategy.

Our findings suggest several open problems in network monitoring that should be addressed in future research. First, monitoring methods need to account for the temporal relationships between graphs in a dynamic network. Furthermore, new methods need to account for the uncertainty associated with community detection and other exploratory methods on networks that are relied upon in network monitoring strategies. Such analyses are not straightforward, but handling uncertainty in estimation would promote the reliability and robustness of the method, as briefly demonstrated in our simulation study. Finally, a better understanding of model misspecification of model-based monitoring approaches is needed. Such a study would illuminate the robustness of a chosen model-based strategy and when nonparametric methods are preferable. For each of these future research avenues and others that are not mentioned, a performance analysis like that done in this manuscript is essential to fully understand the utility of the new monitoring method.

**Conflicts of Interest:** The authors have nothing to disclose.

# Tables

|       | Change                  | Description                                               |
|-------|-------------------------|-----------------------------------------------------------|
| (i)   | $c \to c^*$             | Structural change                                         |
| (ii)  | $P \to P^*$             | Communication outbreaks and change in community structures |
| (iii) | $\theta \to \theta^*$   | Propensities of nodes change                              |

**Table 1.** A description of changes available to the dynamic DCSBM.

| Class of method | Abbreviation | Brief description | No. |
|---|---|---|---|
| Method based on global network summary statistics | | **EWMA chart based on** | |
| | avg. degree | Averaged degree over all nodes | 1 |
| | avg. eigenvector | Averaged eigenvector over all nodes | 2 |
| | avg. betweenness | Averaged betweenness over all nodes | 3 |
| | max betweenness | Maximum betweenness over all nodes | 4 |
| | avg. closeness | Averaged closeness over all nodes | 5 |
| | max.closeness | Maximum closeness over all nodes | 6 |
| | global.cluster.coeff | Global clustering coefficient | 7 |
| | avg.local.cluster.coeff | Averaged local clustering coefficient over all nodes | 8 |
| | diameter | diameter (the largest geodesic distance) | 9 |
| | avg.shortest.path | shortest path length averaged over all node-pairs | 10 |
| | max.shortest.path | maximum shortest path averaged over all nodes | 11 |
| | assortativity | Assortativity | 12 |
| Scan-based | Priebe | Moving window-based scan satictics | 13 |
| Model-based | Wilson | Monitor $P$ using Shewhart charts | 14 |
| | Yu | Monitor $\theta$ using compositional $T^2$ chart | 15 |

**Table 2.** Overview of the 15 monitoring methodologies considered in the comparison.

| Type of statistics | Metrics ($S_t$) |
|---|---|
| Centrality | Average degree |
| | Average eigenvector |
| | Average betweenness, maximum betweenness |
| | Average closeness, maximum closeness |
| Clustering | Global clustering coefficient |
| | Average local clustering coefficient (to the nodes) |
| Distance | Diameter |
| | Average shortest path distance (to a pair of nodes) |
| | Maximum shortest path distance (to the nodes) |
| Homophily | Assortativity |

**Table 3.** Global network summary statistics.

| | | | Monitoring method | | | | | | | | | | | | | | | Best method |
|---|---|---|---|---|---|---|---|---|---|---|---|---|---|---|---|---|---|---|
| | | | Global network summary statistics | | | | | | | | | | | | Scan | Model | | |
| | Scenario | Figure | 1 | 2 | 3 | 4 | 5 | 6 | 7 | 8 | 9 | 10 | 11 | 12 | 13 | 14 | 15 | |
| One-community DCSBM | Global | 5 | ✓✓ | | ✓✓ | ✓ | | | ✓✓ | ✓ | ✓✓ | ✓ | ✗ | ✓ | ✗ | ✓✓ | ✗ | many |
| | Local | 6 | ✓ | | | ✗ | ✗ | ✗ | | | | | ✗ | ✗ | ✗ | | ✗ | 1 |
| | Propensity | 7 | | | | ✓ | ✓ | ✓ | ✓ | | | | ✗ | | | | ✓✓ | 15 |
| Two-community DCSBM $\bar{N}=(10,90)n$ | Intensified | 11 | | | | | | | | | | | | | | ✓ | | 14 |
| | Split | | ✓✓ | | ✓✓ | ✓ | | | ✓✓ | ✓ | ✓ | ✓✓ | ✓ | | | ✓✓ | | many |
| | Merge | | ✓✓ | | ✓✓ | ✓ | | | ✓✓ | ✓ | ✓ | ✓✓ | ✓ | | | ✓✓ | ✗ | many |
| | Form | | ✓✓ | ✓✓ | ✓✓ | ✓✓ | ✓✓ | ✓✓ | ✓✓ | ✓✓ | ✓✓ | ✓✓ | ✓✓ | | ✓✓ | ✓✓ | | many |
| | Fragment | 12 | ✓✓ | ✓✓ | ✓✓ | ✓✓ | ✓✓ | ✓✓ | ✓✓ | ✓✓ | ✓✓ | ✓✓ | ✓✓ | ✓ | | ✓✓ | | many |
| Two-community DCSBM $\bar{N}=(50,50)n$ | Intensified | 8 | ✓✓ | | ✓ | | ✗ | ✗ | ✓ | | | ✓ | ✓ | | ✗ | ✓✓ | ✗ | 1,14 |
| | Split | | | | ✓ | | ✗ | ✗ | | | | ✓ | | ✗ | ✗ | ✓✓ | ✗ | 14 |
| | Merge | 9 | | | ✓ | ✗ | ✗ | ✗ | | | | ✓ | | ✗ | ✗ | ✓✓ | ✗ | 14 |
| | Form | | ✓✓ | ✓ | ✓ | | ✗ | ✗ | ✓✓ | ✓ | | ✓ | ✓ | | | ✓✓ | ✗ | 1, 7, 14 |
| | Fragment | 10 | ✓✓ | ✓ | ✓ | | | | ✓ | ✓ | | ✓ | ✓ | ✗ | | ✓✓ | ✗ | 1, 14 |

**Table 4**: Overview of the performance of monitoring methods. Where ✗ indicates poor performance, ✓ indicates moderate performance and ✓✓ indicates good performance.

# Figures

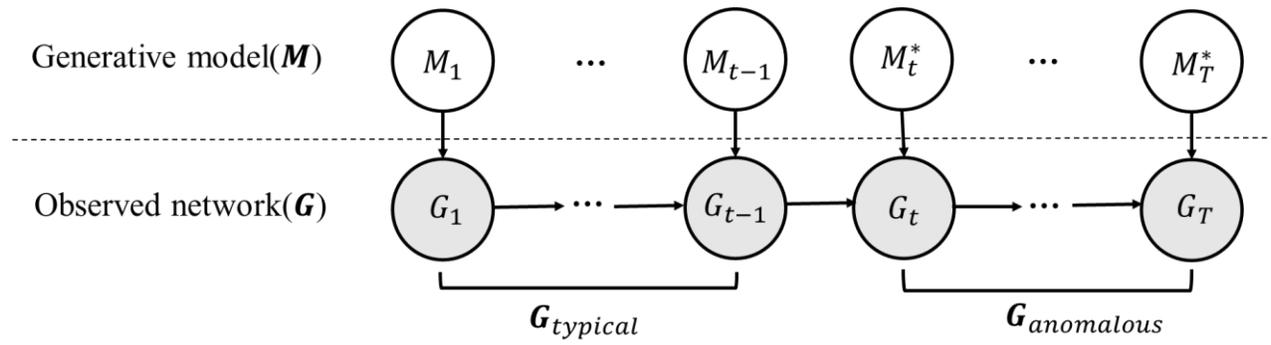

**Figure 1**. The dynamic network generative process. $M_t$'s are generative models and $G_t$'s are the resulting observed networks. Change is introduced at $t^*$ where $M_{t:t<t^*} \neq M_{t:t\geq t^*}$.

| | Change description | Graphic representation | Parameter change representation |
|---|---|---|---|
| **One Community DCSBM** | | | $p \to p^*$ |
| | Global change | 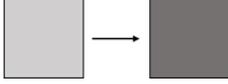 | $0.2 \to 0.25$ |
| | Local change | 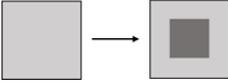 | $0.2 \to 0.4$ for $n/5$ |
| | | | $\boldsymbol{\theta} \to \boldsymbol{\theta}^*$ |
| | Propensity change | 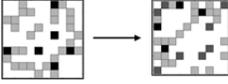 | $U(0.5, 1.5) \to U(0.5, 3)$ |
| **Two Communities DCSBM** | | | $\boldsymbol{P} \to \boldsymbol{P}^*$ |
| | Intensified communication | 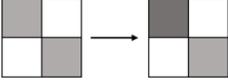 | $\begin{pmatrix} 0.3 & 0.1 \\ 0.1 & 0.3 \end{pmatrix} \to \begin{pmatrix} 0.4 & 0.1 \\ 0.1 & 0.3 \end{pmatrix}$ |
| | | | $\boldsymbol{c} \to \boldsymbol{c}^*$ |
| | | | *Achieved by change on $\boldsymbol{P} \to \boldsymbol{P}^*$* |
| | Split | 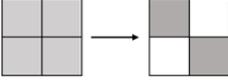 | $\begin{pmatrix} 0.2 & 0.2 \\ 0.2 & 0.2 \end{pmatrix} \to \begin{pmatrix} 0.3 & 0.1 \\ 0.1 & 0.3 \end{pmatrix}$ |
| | Merge | 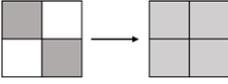 | $\begin{pmatrix} 0.3 & 0.1 \\ 0.1 & 0.3 \end{pmatrix} \to \begin{pmatrix} 0.2 & 0.2 \\ 0.2 & 0.2 \end{pmatrix}$ |
| | Form | 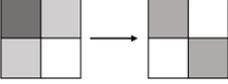 | $\begin{pmatrix} 0.4 & 0.2 \\ 0.2 & 0.1 \end{pmatrix} \to \begin{pmatrix} 0.3 & 0.1 \\ 0.1 & 0.3 \end{pmatrix}$ |
| | Fragment | 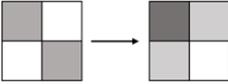 | $\begin{pmatrix} 0.3 & 0.1 \\ 0.1 & 0.3 \end{pmatrix} \to \begin{pmatrix} 0.4 & 0.2 \\ 0.2 & 0.1 \end{pmatrix}$ |

**Figure 2**. Taxonomy and representation of network change when $n = 40$.

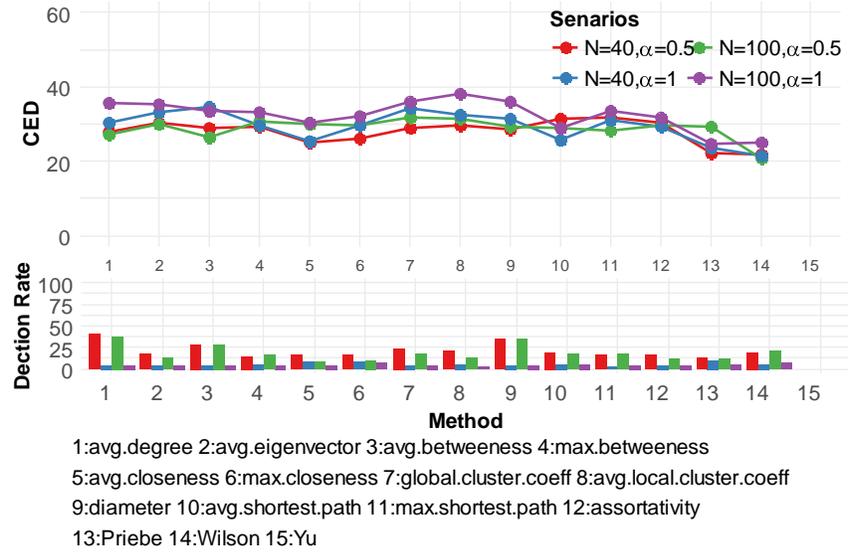

**Figure 3:** Comparison of methods on the *no-change* scenario with one community DCSBM.

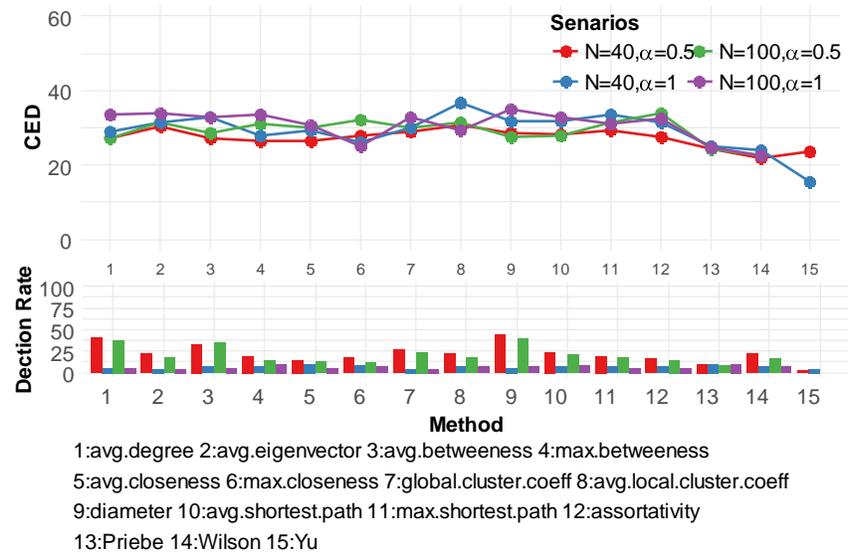

**Figure 4:** Comparison of methods on the *no-change* scenario with two-community DCSBM.

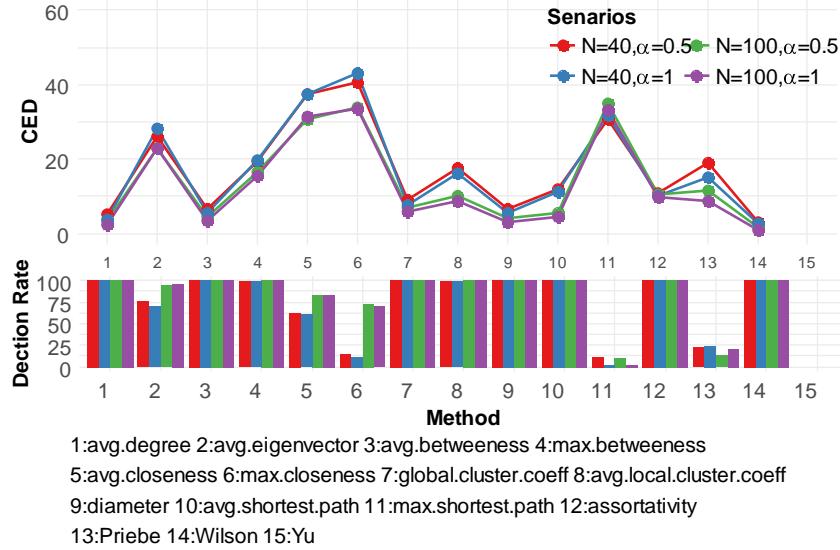

**Figure 5:** Comparison of methods on the *global* change scenario with one-community DCSBM.

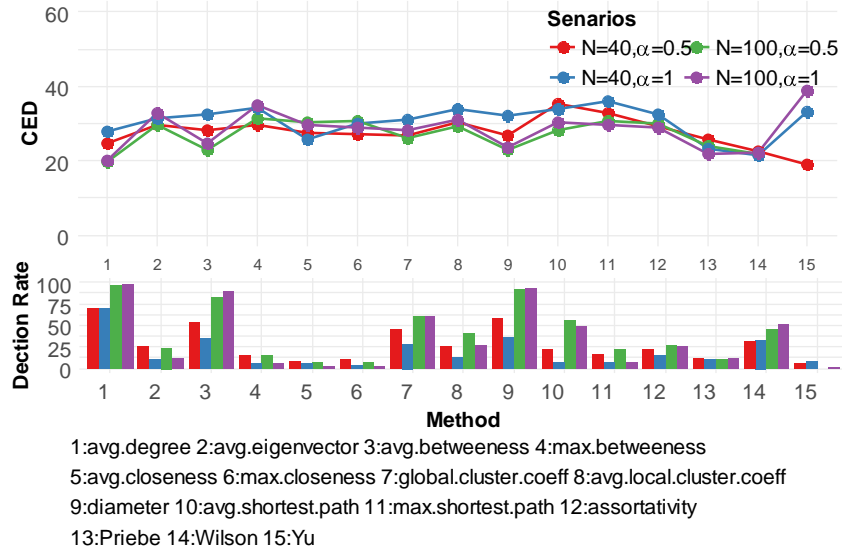

**Figure 6:** Comparison of methods on the *local* change scenario with one-community DCSBM.

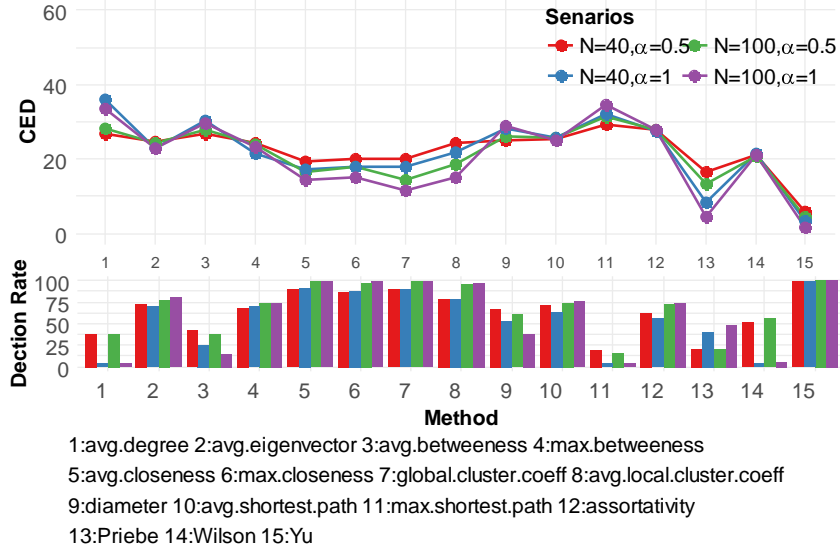

**Figure 7:** Comparison of methods on the *propensity* change scenario with one-community.

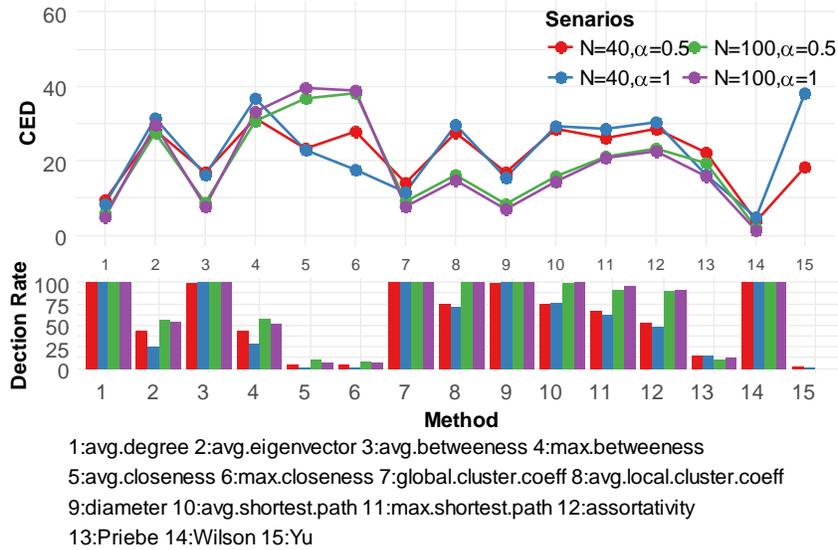

**Figure 8.** Comparison of methods on the *intensified communication* scenario with two-community DCSBM.

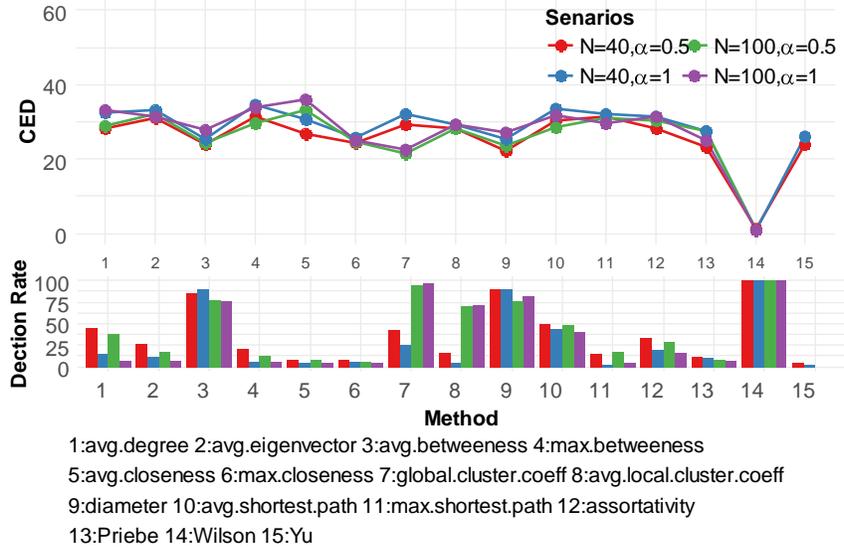

**Figure 9:** Comparison of methods on the *merge* scenario with two-community DCSBM.

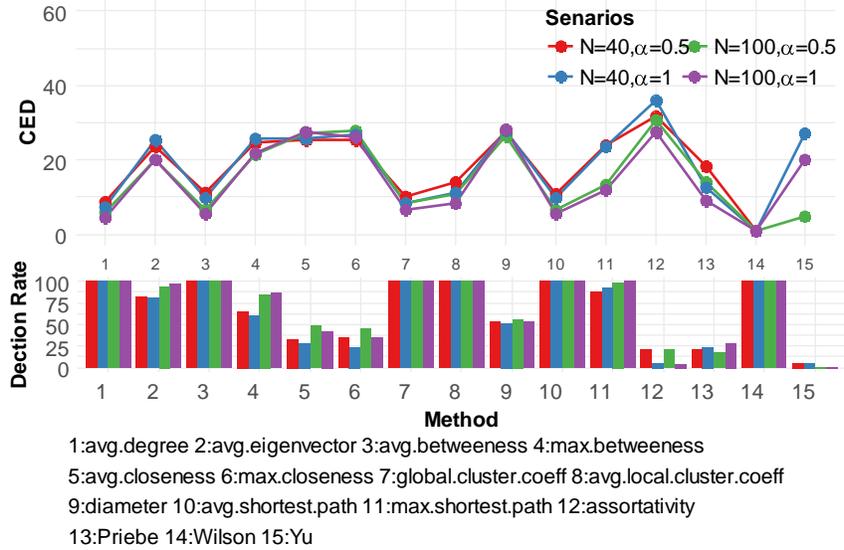

**Figure 10:** Comparison of methods on the *fragment* scenario with two-community DCSBM.

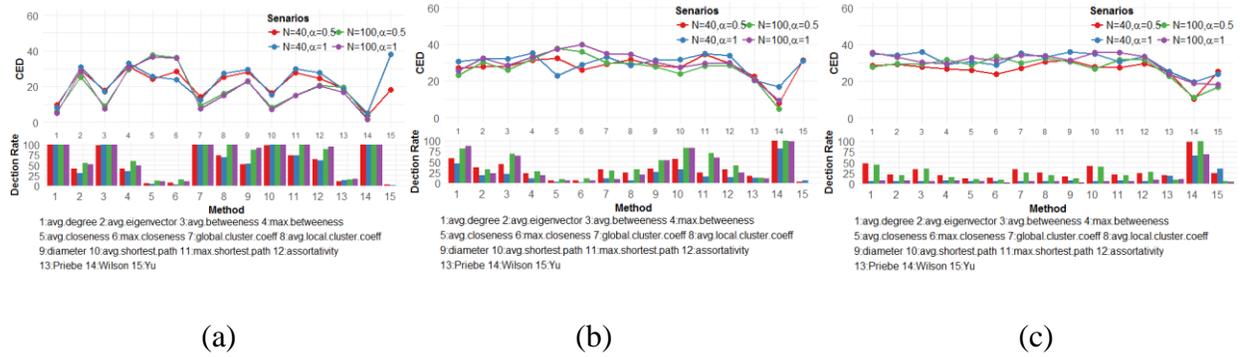

(a)                  (b)                  (c)

**Figure 11:** Effect of node configuration on the comparison of methods in the *intensified communication* scenario with two-community DCSBM. Size of community with increased communication levels: (a) 50%, (b) 25%, and (c) 10%.

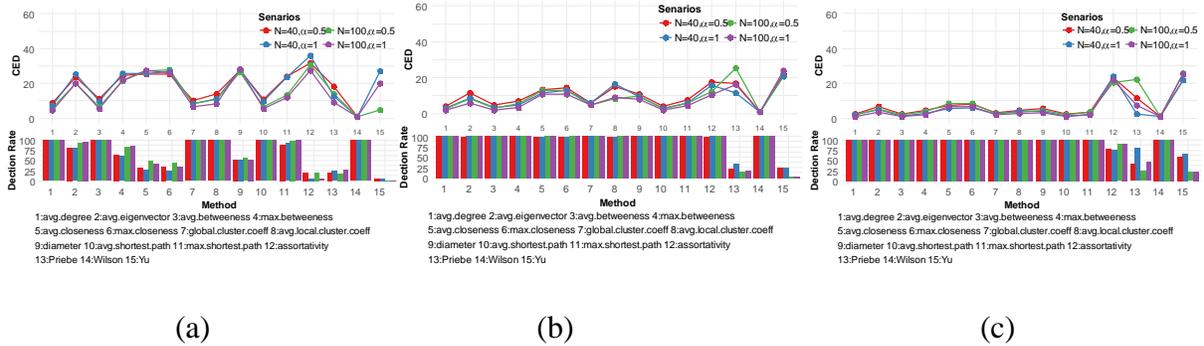

(a)                  (b)                  (c)

**Figure 12:** Effect of node configuration on the comparison of methods in the *fragment* scenario with two-community DCSBM. Relative community size changes: (a) 50%, (b) 25%, and (c) 10%.

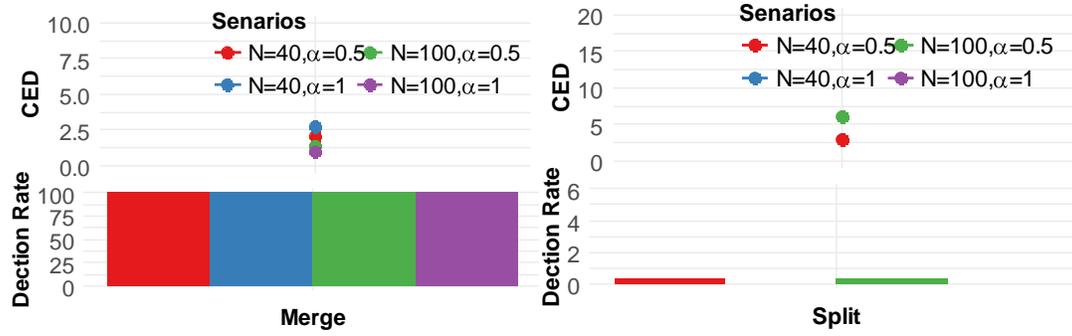

**Figure 13**: The performance of Wilson's method on *merge* and *split* scenarios in two-unknown-community DCSBM.